\documentstyle[fleqn,11pt,epsf]{article} 

\newcommand{\erf}{\mathop{\rm erf}\nolimits}%
\newcommand{\sgn}{\mathop{\rm sgn}\nolimits}%
\renewcommand{\baselinestretch}{1.5}

\begin{document}
\vspace{-3cm}
\begin{minipage}[t]{3cm}
  \vspace{0pt}
  {PREPRINT}\\
  {July 20, 1994}
\end{minipage}
\hfill
\begin{minipage}[t]{3cm}
  Univ. zu K\"{o}ln\\
  UvA/ITF 16/94
\end{minipage}
\vspace{8mm}

\noindent
%{\large {\bf version date:} July 20, 1994} \\[5mm]
\noindent
\begin{center}
{\Large\bf Existence of an incommensurate ground state
 of interacting spinless fermions in infinite dimensions} \\[6mm]
 {\Large G.S. UHRIG${}^1$ and R. VLAMING${}^2$}
 \\[3mm]
 {${}^1$ \large \em  Inst. f. Theor. Physik,
  Univ. zu K\"oln,
  50937 K\"oln, Germany}
 \\
 {${}^2$ \large \em 
Inst. v. theor. fysica, UvA, 1018 XE, Amsterdam, The Netherlands}
 \\
\end{center}

\renewcommand{\baselinestretch}{1.5} \small \normalsize
\begin{abstract}
%\normalsize
\noindent
We focus on the ground state phase diagram of
a system of spinless fermions with repulsion on a hypercubic
lattice in the limit of infinite dimensions.
It occurs spontaneous symmetry breaking into a charge
density wave (CDW).
Using an ansatz for the order parameter which includes the
homogeneous, the AB- and a large class of incommensurate 
phases we are able to calculate the phase diagram.
\end{abstract}

%\noindent
%{\bf Keywords:} spinless fermions, incommensurate states, infinite 
%dimensions. \\
%{{\bf Address for correspondation:} \\
%Dr. R. Vlaming, instituut voor theoretische fysica \\
%Universiteit van Amsterdam, 1018 XE Amsterdam, The Netherlands \\
%tel: ++31-20-5255771, fax: ++31-20-5255778, email: vlaming@phys.uva.nl \\ }
%\newpage

\renewcommand{\baselinestretch}{1.3} \small \normalsize
\vspace{5mm}
It is an important object to disclose the structure of the phase diagram 
of models  with strong electron correlation. The problem is twofold.
First, one
has to have a clear picture which phases are likely
to be realised and second,
for these phases it must be possible to calculate
the free energy. In case of the Hubbard model for example
{both} problems can hardly be surmounted. Even if one restricts oneself
to a small number of phases  and to the dynamic mean field theory 
resulting from
the limit of high dimensions \cite{vollh93}, the problem still requires 
massive computational effort (e.g.\ \cite{jarre92}). 
The situation is different for the model of interacting
spinless fermions \cite{yang66undmehr} %,cloiz66,cloiz66b,lembe93,shank94} 
on an infinite dimensional hypercubic
lattice. Although the diagrammatic theory is simple, 
the physics is far from trivial. 

The Hamiltonian  of the
 spinless fermion model in second quantisation is
\begin{equation}
\hat H = -\frac{t}{\sqrt{2d}}\sum\limits_{<i,j>}
\hat c_i^+ \hat c_j ^{\phantom{+}}
+ \frac{U}{4d}\sum\limits_{<i,j>} \hat n_i \hat n_j -\mu \sum
\limits_{i} \hat n_i   \ .
\end{equation}
Scaling with the inverse dimension $1/d$ is performed 
 to ensure the continuity
 of the limit of infinite dimensions \cite{vollh93,metzn89,mulle89}. 

In a previous work the stability of the homogeneous phase was investigated
in the limit $d\to\infty$ \cite{uhrig93b}. 
In this limit
Hartree- and random phase approximation become exact 
\cite{mulle89,uhrig93b}.
Besides the occurrence of an  AB-charge density
wave (CDW) characterised by the wave vector 
$\mbox{\boldmath$Q$}:=(\pi,\pi,\ldots,\pi)^\dagger$
it was found that the density-density susceptibility 
$\chi(\mbox{\boldmath$q$})=\left(\chi^{-1}_0(\mbox{\boldmath$q$})
+U\eta(\mbox{\boldmath$q$})\right)^{-1}$ 
diverges at
incommensurate values of the wave vector $\mbox{\boldmath$q$}$. 
The parameter $\eta$ is defined as
$\eta(\mbox{\boldmath$q$}) := d^{-1}\sum_{i=1}^d \cos (q_i)$
and $\chi^{-1}_0(\mbox{\boldmath$q$})$ is the
 inverse bare susceptibility. 
Commensurate values of $\mbox{\boldmath$q$}$ which belong to a value of
 $\eta\neq -1, 1$ are of measure
zero. The complete stability analysis is possible 
in $d=\infty$ because $\chi_0$
depends on $\mbox{\boldmath$q$}$ only via $\eta$
\cite{mulle89}. Thus it is sufficient to examine $\chi(\eta)$
\cite{freer93}.

The divergence of $\chi(\eta)$ indicates
a second order transition from the homogeneous phase at high doping and/or
low interaction into an incommensurate CDW. 
The structure of the new incommensurate phase has not yet been
investigated. In finite dimensions and for related models such as the Hubbard model
and the $t$-$J$ model 
there exist many works which treat different variational or approximative
approaches to incommensurate
phases \cite{eder91undmehr}. %,shrai89,kane90,jayap89,dzier92,dzier93}. 
%BEGIN{AENDERUNG}
Since the perturbation theory of the model (1) becomes exactly tractable
in $d\to\infty$ it is the natural candidate to examine the energetic
effects of different spatial structures of the order parameter. 
The resulting equations are self-consistent and include infinitely
many parameters in the thermodynamic limit so that no systematic
approach to their solution exists \cite{uhrig93b}.
%Since the spinless fermion model in $d\to \infty$ 
%is exactly solvable on the 
%diagrammatic level it is the natural candidate to examine the energetic
%effects of different spatial structures of the order parameter.
%END{AENDERUNG}
Here we give results for the following ansatz  for the 
spatial structure of the order parameter $b(\mbox{\boldmath$r$}):=
 \langle\hat n_{\mbox{\boldmath$r$}}\rangle
- n$ where $n$ is the average particle density:
$b(\mbox{\boldmath$r$})= b_0 \prod_{i=1}^{d} u_i(r_i)$.
The index $i$ counts the directions; $r_i$ is the 
$i$th component of
{\boldmath$r$}, 
%BEGIN{AENDERUNG}
$b_0$ is the overall amplitude and the new defined functions 
%END{AENDERUNG}
$u_i$ take the values $\pm1$. 
The product form of this
ansatz allows to profit best from the simplifications of 
$d\to\infty$. Yet it is still very general since it allows any  
sequence of $+1$ and $-1$
independently for the $d$ directions without assuming translational
invariance. 

A close inspection of the free energy 
shows that it depends
on $b(\mbox{\boldmath$r$})$ only via the amplitude $b_0$
 and the relative frequency $h$ that
$b(\mbox{\boldmath$r$})$ has the same sign on adjacent sites
\cite{uhrig94b}.
These are the only parameters which are relevant for the 
influence of the
spatial structure on the energy. In order to find an 
explicit functional for the
energy we exploit the freedom in choosing the $u_i$. 
For $i\le hd$ we set $u_i(r_i)\equiv 1$ and for
$i > hd$ we set $u_i(r_i)\equiv (-1)^{r_i}$.
This choice is special since
the order parameter can be characterised by the wave vector 
\begin{equation}
\mbox{\boldmath$Q$}_h=(\underbrace{0,0,\ldots,0}_{hd},
\underbrace{\pi,\pi,...,\pi}_{(1-h)d})^\dagger
\end{equation}
%BEGIN{AENDERUNG}
but does not restrict the freedom contained in 
the ansatz for $b(\mbox{\boldmath$r$})$.
This gives a general relationship $\eta=2h-1$, 
although only for special
realisations $\eta$ can be defined.
%END{AENDERUNG}
The easy structure of the CDW with wave vector $\mbox{\boldmath$Q$}_h$
 helps to resum the perturbation series for the grand canonical potential
since the Green function is a $2\times2$ matrix in $\mbox{\boldmath$k$}$-space.

In infinite dimensions the diagram for the free energy is simple.
Compared to the free fermion system the grand potential
gains a term $\Delta\Omega=T\,\mbox{Tr}\int_0^1d\lambda\,\lambda^{-1}\,
G\Sigma|_{U\rightarrow U\lambda}$, where $G$ is the Green function
and $\Sigma$ the self-energy. At $T=0$ this
yields the ground state energy
\begin{eqnarray}
%\nonumber
  E(\Delta,\eta)\!\!&=&\!\!\frac{Un^2}{2} 
  - \frac{\Delta^2}{2\eta U} +
  \!\int\limits_{-\infty}^{\infty}\!\! dv\;
  \frac{\exp\left(-\frac{v^2}{1-\eta}\right)}{2\sqrt{\pi(1-\eta)}} 
%  \cdot \\  & &\;\;\;
  \left\{\frac{\sqrt{v^2+\Delta^2}}{\sgn(v)}
  \,\erf(P)
  -\frac{\sqrt{1+\eta}}{\sqrt{\pi}}
  \exp(-P^2)\right\}  
\\ \nonumber
 P&:=&\frac{\tilde{\mu}-\sgn(v)\sqrt{v^2+\Delta^2}}
  {\sqrt{1+\eta}}
\end{eqnarray}
%BEGIN{AENDERUNG}
where $\tilde{\mu}=\mu-nU$ and $\Delta=-\eta U b_0$. The doping $\delta=0.5-n$ is given by
\begin{equation}
\delta=\int\limits_{-\infty}^{\infty}\!\! dv\;
  \frac{\exp\left(-\frac{v^2}{1-\eta}\right)}{2\sqrt{\pi(1-\eta)}}\,
  \erf(P)
\end{equation}
and the equations which define the energetic minimum are
$\partial E/\partial \Delta|_n=0$ and 
$\partial E/\partial \eta|_n=0$.
%END{AENDERUNG}
For a particular value of the interaction $U$ it is
possible to determine for each doping the parameters
$\Delta$ and $\eta$ which describe the physical state.
Additionally, one has to be on the watch
for  first order phase transitions.

%%%%INSERT FIGURE 1 DIRECTLY ABOVE OR BELOW THIS PARAGRAPH %%%%%
As example we look at the case $U=1.2$. We plotted ground state energy
 as function of doping in Fig.\ 1. 
For large doping and at half-filling the situation is clear. For the
former the system is in the homogeneous phase whereas for the 
latter the system is in the commensurate CDW
characterised by $\eta=-1$ (AB-phase). Inbetween four special 
points at 
$\delta_{\mbox{\tiny div-AB}}$, 
$\delta_{\mbox{\tiny PS-IP}}$, 
$\delta_{\mbox{\tiny  PS-AB}}$, and 
$\delta_{\mbox{\tiny div-IP}}$ 
can be identified.  

Let us focus on the situation with fixed $\eta=-1$. The system is forced
to choose between the homogeneous phase and the AB-phase. 
 The response
of the system towards fluctuations with wave vector $\mbox{\boldmath$Q$}_0$ 
becomes infinite at  $\delta_{\mbox{\tiny div-AB}}$. The 
ground state energy in indicated by a dotted and a dotted-dashed
curve, respectively in Fig.\ 1. For 
smaller doping the ground state energy is concave so that a
first order phase transition will occur. The Maxwell construction
gives rise to the value $\delta_{\mbox{\tiny  PS-AB}}$. 
Thus, there is phase separation between the homogeneous phase
and the AB phase for $0<\delta<\delta_{\mbox{\tiny  PS-AB}}$.
This situation turns out to be generic for 
$U<U_{\mbox{\tiny IPL}}\simeq 0.5716$ and 
$U>U_{\mbox{\tiny IPH}}\simeq 1.9145$. Inbetween -- and
thus for the case $U=1.2$ -- a different  scenario
is realised. 

The divergence of $\chi(\eta)$ is at $\delta_{\mbox{\tiny div-IP}}$
for general values of $\eta$.
For $U_{\mbox{\tiny IPL}}<U<U_{\mbox{\tiny IPH}}$ holds
$\delta_{\mbox{\tiny div-IP}} > \delta_{\mbox{\tiny PS-AB}}$.
By minimizing $E(\Delta,\eta)$ we obtain
the solid curve in Fig.\ 1. 
 Again, the curve is concave for low dopings 
so that a Maxwell construction
must be made. The doping where the phase separation sets in is 
$\delta_{\mbox{\tiny PS-IP}}$. It is important to realise that 
$\delta_{\mbox{\tiny PS-IP}} < \delta_{\mbox{\tiny PS-AB}}$.
For dopings $\delta_{\mbox{\tiny PS-IP}} < \delta <
\delta_{\mbox{\tiny div-IP}}$ the 
pure incommensurate phase is the absolute energy minimum. 

%%%%INSERT FIGURE 2 DIRECTLY ABOVE OR BELOW THIS PARAGRAPH %%%%%
The overall result is depicted in 
Fig.\ 2. At half-filling the system is in the AB-phase for 
all $U>0$. For large doping the system is in the homogeneous
phase. For $U<U_{\mbox{\tiny IPL}}$ and 
$U>U_{\mbox{\tiny IPH}}$ there is phase separation between the
AB-phase and a hole-enriched homogeneous phase.
For $U_{\mbox{\tiny IPL}}<U<U_{\mbox{\tiny IPH}}$ there exists
a region where an incommensurate phase is present,
 and a region of phase separation between the
 AB-phase and the hole-enriched incommensurate phase.

%%%%INSERT FIGURE 3 DIRECTLY ABOVE OR BELOW THIS PARAGRAPH %%%%%
The parameter $\eta$ is close to $-1$ in the
incommensurate phase. To give an impression we plotted those
values of $\eta$ in Fig.\ 3 
which arise at the borders of the incommensurate phase in Fig.\ 2.
The function $\eta(\delta)$ at constant $U$ is monotonic so
that the
figure gives a good impression of the deviation of $\eta$ from $-1$.
Note that the values for $\eta$ belonging to the left border (Fig.\ 2),
given by the solid curve in Fig.\ 3,
are important for the whole region of phase separation.

In summary, we demonstrated the complexity of the ground state 
phase diagram of the model of interacting spinless fermions
in infinite dimensions.
Most important is the existence of an incommensurate 
phase and of the phase separation between the AB-phase and
the incommensurate or homogeneous phase.

This work was supported by the Deutsche 
Forschungsgemeinschaft (SFB 341)
and the Stichting voor Fundamenteel Onderzoek der Materie (FOM).

%\newpage
%\bibliographystyle{physrev}
%\bibliography{liter7a,liter7b}

\section*{Figure captions}
\begin{quote}
{\bf Figure 1.}
Ground state energy for $U=1.2$. Dotted curve: 
homogeneous phase. Dotted-dashed curve: AB phase.
Solid curve: incommensurate phase ($\eta$ optimised). 
Divergence of $\chi(-1)$ at $\delta_{\mbox{\tiny div-AB}}$.
Divergence of $\chi(\eta)$ for optimised $\eta$ 
at $\delta_{\mbox{\tiny div-IP}}$.
Maxwell constructions between $\delta=0$ and 
$\delta_{\mbox{\tiny PS-AB}}$ or 
$\delta_{\mbox{\tiny PS-IP}}$. 
Incommensurate phase between $\delta_{\mbox{\tiny PS-IP}}$
and $\delta_{\mbox{\tiny div-IP}}$.
To facilitate
visibility a linear function of $\delta$ has been added:
 $a\simeq0.5561$, $b\simeq0.1946$.

{\bf Figure 2.}       
Phase diagram. HOM: homogeous phase, AB: AB phase, 
IP: incommensurate phase, PS-AB:
phase separated region between homogeneous and AB phase, PS-IP:
phase separated region between incommensurate and AB phase.
$U_{\mbox{\tiny IPL}}$, $U_{\mbox{\tiny IPH}}$: lower and 
upper interaction bound of the IP.

{\bf Figure 3.}
The value of $\eta$ along the phase borders of the incommensurate
phase. Solid curve: right border; dashed curve: left border.

\end{quote}

%\newpage
\begin{figure}[h]
\centerline{\epsffile{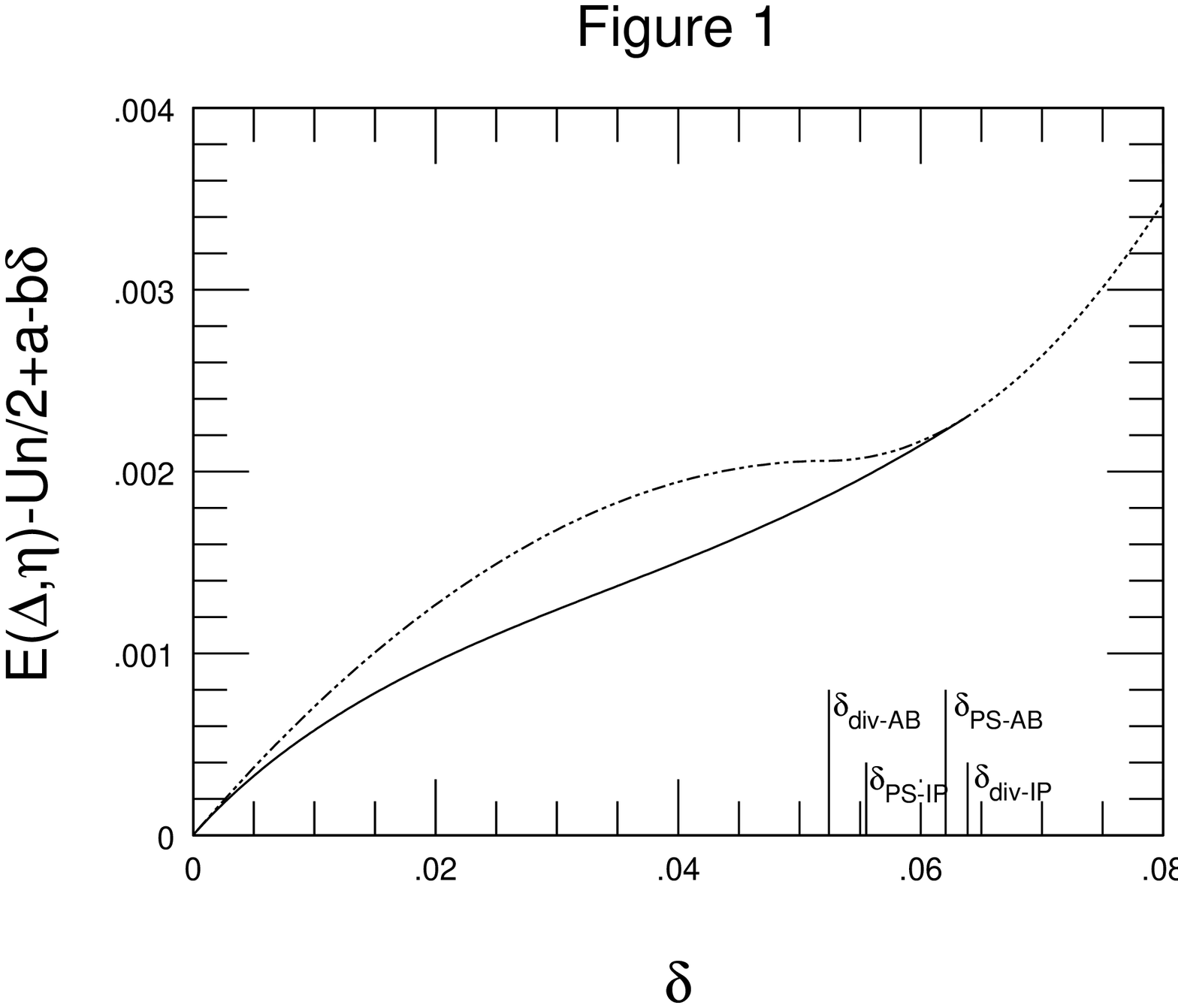}}
\end{figure}

%\newpage
\begin{figure}[h]
\centerline{\epsffile{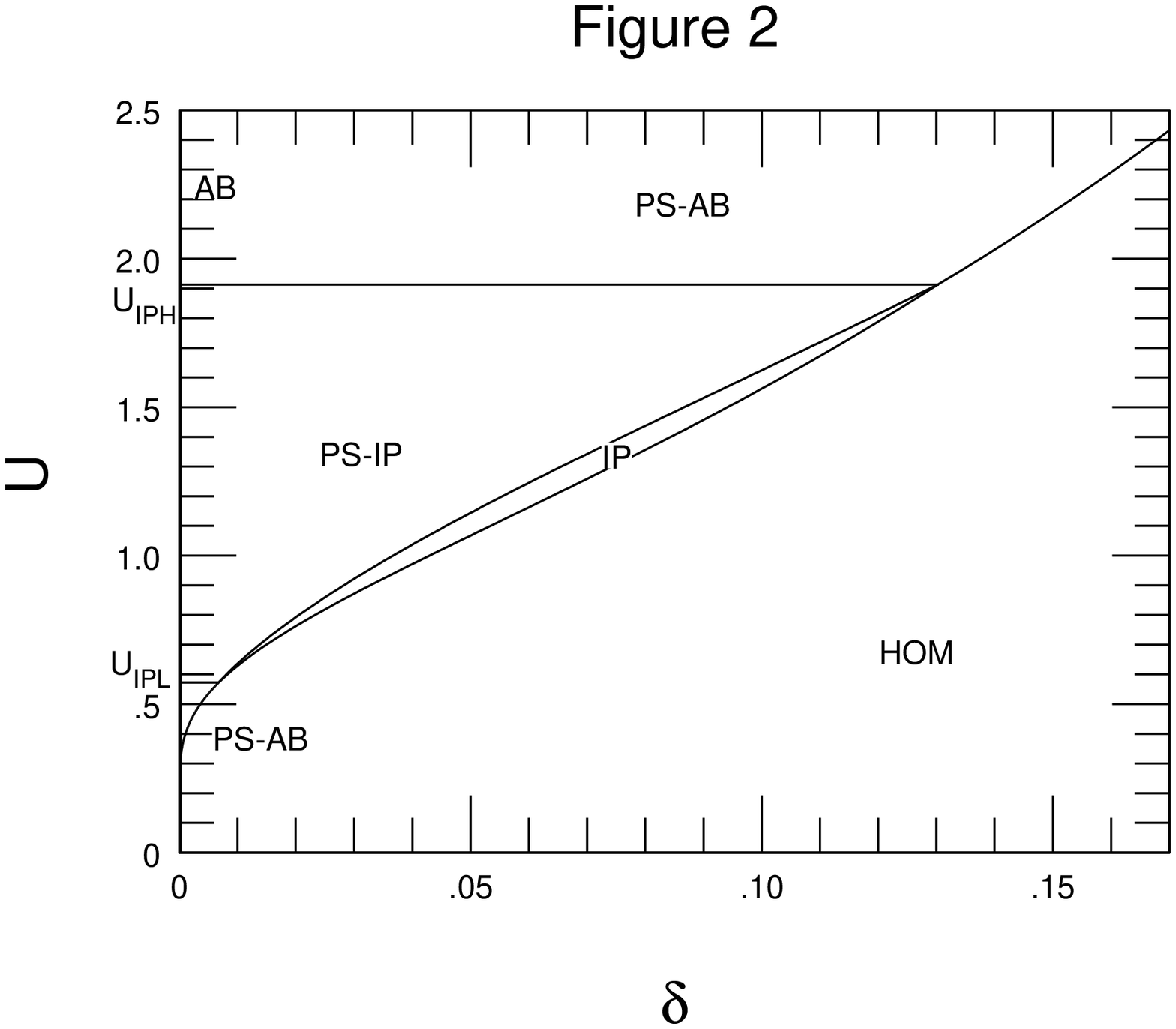}}
\end{figure}

%\newpage
\begin{figure}[h]
\centerline{\epsffile{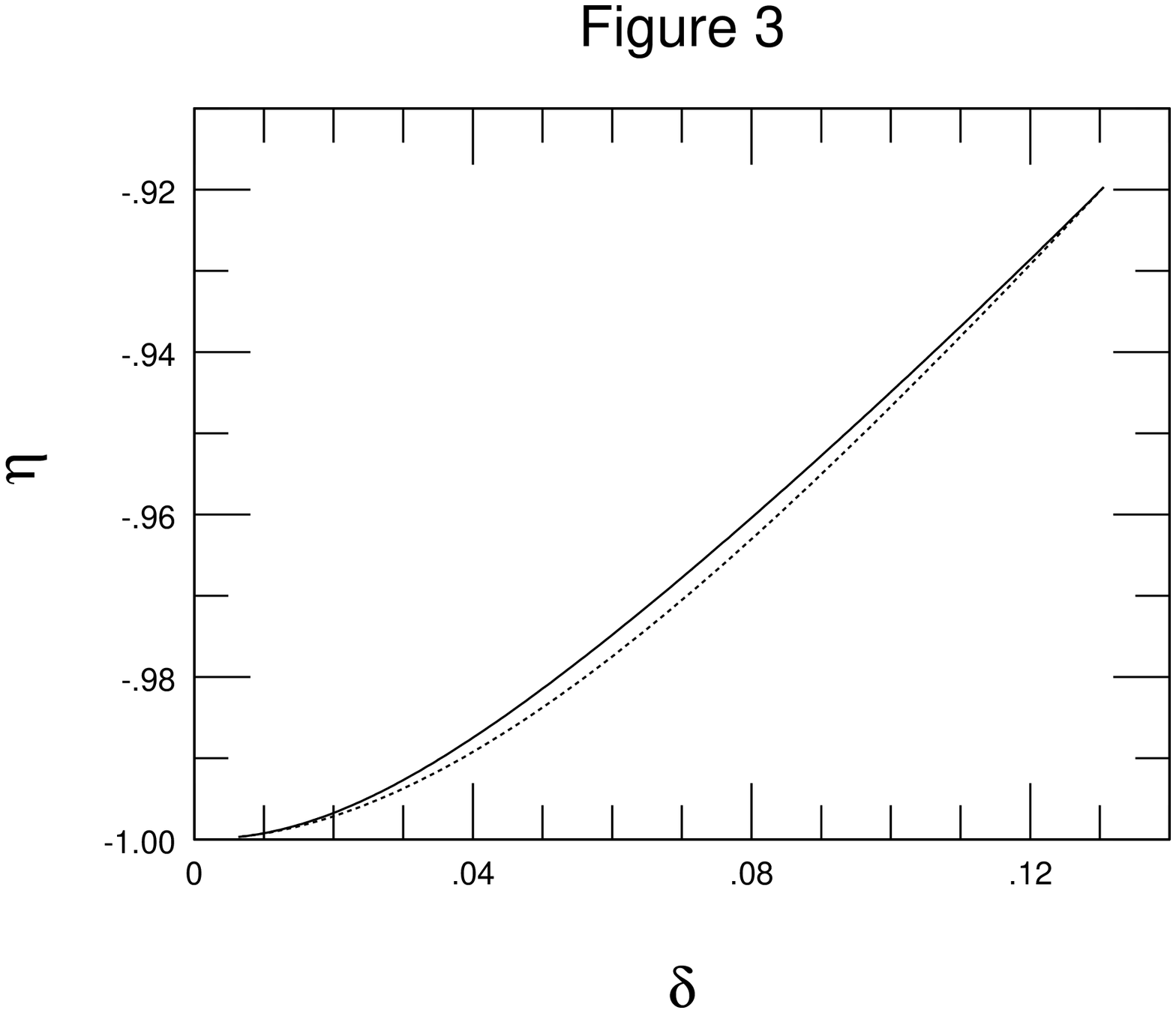}}
\end{figure}

\end{document}